\documentclass[twocolumn,showpacs,amsmath,nofootinbib,pra,aps,amssymb,longbibliography]{revtex4-1}

\usepackage[T1]{fontenc}
\usepackage[utf8]{inputenc}
\usepackage{times}
\usepackage{color} 
\usepackage{array}
\usepackage{amssymb,amsmath}
\usepackage{amsbsy}
\usepackage[pdftex]{graphicx} 
\usepackage{bm}
\usepackage{float}
\usepackage{dcolumn}
 
\usepackage[unicode,breaklinks]{hyperref}
\hypersetup{
    unicode=true,
    plainpages=false, 
    colorlinks=true,
    linkcolor=blue,
    citecolor=blue,
    filecolor=black,
    urlcolor=blue
}
\usepackage{url}
\usepackage{verbatim}

\newcommand{\bk}{\mathbf{k}}
\newcommand{\br}{\mathbf{r}}

\newcommand{\bx}{\mathbf{x}}

\newcommand{\Ut}{\tilde{U}}

\begin{document}

\title{Quantum droplet states of a binary magnetic gas}
  \author{Joseph~C.~Smith, D.~Baillie, and P.~B.~Blakie   }
	
	\affiliation{Dodd-Walls Centre for Photonic and Quantum Technologies, New Zealand}
	\affiliation{Department of Physics, University of Otago, Dunedin 9016, New Zealand}
  
\date{\today}
\begin{abstract}  

Quantum droplets can emerge in bosonic binary magnetic gases (BMGs) from the interplay of short- and long-ranged interactions, and quantum fluctuations. We develop an extended meanfield theory for this system and use it to predict equilibrium and dynamical properties of BMG droplets.  We present a phase diagram and characterize miscible and immiscible droplet states. We also show that a single component self-bound droplet can be used to bind another magnetic component which is not in the droplet regime.   Our results should be realizable in experiments   with mixtures of highly-magnetic lanthanide atoms.
\end{abstract}
\maketitle

\noindent\textbf{Introduction} -
Quantum droplets are a dilute liquid-like state of an ultra-cold atomic gas that occurs when collapse-inducing attractive two-body interactions are balanced by a repulsive term with a higher-order dependence on the density \cite{Bulgac2002a}. 
Experimental studies over the past four years have produced quantum droplets in two different bosonic systems in which the collapse is stabilized by quantum fluctuations\footnote{Here the  energy  of the two-body interactions and quantum fluctuations scale with the density $n$ as $\mathcal{E}_\text{2B}\sim n^2$ and $\mathcal{E}_\text{QF}\sim n^{5/2}$, respectively.}: 
 (i) Binary mixtures of atoms with an attractive interspecies interaction \cite{Petrov2015a}. These droplets have been prepared with mixtures of potassium spin states  \cite{Cabrera2018a,Semeghini2018a}, and with a heteronuclear potassium-rubidium \cite{DErrico2019a} mixture.  (ii) Single-component systems of highly magnetic atoms in the regime where the magnetic dipole-dipole interaction (DDI) dominates over the contact interaction.  Gases of  dysprosium (Dy) \cite{Ferrier-Barbut2016a} and erbium (Er) \cite{Chomaz2016a} atoms have been used to prepare these droplets.

Here we consider a new class of quantum droplet formed in a BMG, i.e.~a two-component bosonic gas, where the atoms of each component  have a large magnetic moment. In this system each component can independently form a droplet, but together the components have an array of interactions that lead to a multi-component droplet with unique features. Critically, interactions can favour the droplet being either in a miscible or immiscible phase. In the latter case the long-ranged DDIs play an important role in organising the phase-separated parts of the droplet. Such multi-component physics is absent from  (non-magnetic) binary mixtures, for which the interspecies attraction only allows droplets to form in the miscible regime, where it effectively behaves as a single component system (see \cite{Petrov2015a}).
Motivation for understanding this new system comes from recent experimental progress with BMGs \cite{Trautmann2018a}, where  binary Bose-Einstein condensates with five different Er-Dy isotope combinations have been produced.

\noindent\textbf{Formalism} -
We consider a zero temperature gas of two  bosonic atoms with a large magnetic moment $\mu^m_i$ ($i=1,2$), polarized along the $z$ axis. In this work we assume both species have the same mass $m$, which well-approximates any mixture of Er and Dy isotopes.
The stationary states of species $i$ are described by the extended GPE $\mathcal{L}_i\psi_i = \mu_i\psi_i$, where
\begin{align}
   \mathcal{L}_i &= h_{\text{sp}} +\sum_jg_{ij}n_j+ \sum_jg_{ij}^\text{dd}\Phi_j(\bx) + \Delta\mu_i,\label{LGPEi}
\end{align}
and $\mu_i$ is the chemical potential of species $i$, used to fixed the number of atoms $N_i=\int  d\bx\,|\psi_i|^2$. Here $h_{\text{sp}}=-\frac{\hbar^2\nabla^2}{2m}+V(\bx)$ is the single-particle Hamiltonian, where $V(\bx)$ represents any external potential.
The short-ranged two-body interaction coupling constants are  $g_{ij} = 4\pi\hbar^2 a_{ij}/m$, where $a_{ij}$ is the $s$-wave scattering length between species $i$ and $j$. The long-ranged DDIs are described by 
%
  $\Phi_i(\bx) =  \int d\bx'\,f^\text{dd}(\bx-\bx')|\psi_i(\bx')|^2$,
where $f^\text{dd}(\br) = \frac{3}{4\pi r^3}(1-3\cos^2\theta)$, with the coupling constant $g_{ij}^\text{dd}=4\pi \hbar^2a_{ij}^\text{dd}/m$ and $a_{ij}^\text{dd}=m\mu_0\mu_i^m\mu_j^m/12\pi\hbar^2$ being the dipole length.   

The quantum fluctuation effects included in the eGPE are described by the terms $\Delta \mu_i$, derived by extending the formalism of Ref.~\cite{Petrov2015a} to include DDIs.
 For a uniform (and equal mass) dipolar mixture of component density $n_i$ the quantum fluctuations contribute an energy density of $\mathcal{E}_{\mathrm{QF}} = \frac{\sqrt2 m^{3/2}}{15\pi^2\hbar^3}  \sum_\pm\int_0^{\pi/2} d\theta_k\,\sin\theta_k I_{E\pm}^{5/2}$ where
\begin{align}
    I_{E\pm} = n_1\Ut_{11}+n_2\Ut_{22}\pm \sqrt{\delta_1^2+4\Ut_{12}^2n_1n_2},
\end{align}
with $\delta_1 = n_1\Ut_{11}-n_2\Ut_{22}$, and  $\Ut_{ij}(\bk) = g_{ij} + g_{ij}^\text{dd}(\cos^2\theta_k-\tfrac13)$ being the Fourier transform of the total interaction potential.
From this we derive the  chemical potential correction as
 $\Delta\mu_i =  \partial_{n_i}\mathcal{E}_{\mathrm{QF}}=\frac{m^{3/2}}{3\sqrt{2}\pi^2\hbar^3}  \sum_\pm\int_0^{\pi/2}  d\theta_k\,\sin\theta_k \,I_{i\pm}$, where
 \begin{align}
I_{1\pm}&=  \Biggl(\Ut_{11} \pm \frac{\delta_1 \Ut_{11}+2\Ut_{12}^2n_{2}}{\sqrt{\delta_1^2+4\Ut_{12}^2n_1n_2}}\Biggr)I_{E\pm}^{3/2}.
\end{align}
  We omit writing the similar expression for $I_2$.  This theory will provide a good description in the dilute regime, which is well-satisfied in ultra-cold atomic experiments. To employ this result in the inhomogeneous situation of Eq.~(\ref{LGPEi}) we make the local density approximation and set $n_i=|\psi_i(\bx)|^2$.  In using these expressions we ignore the small imaginary part of $\Delta \mu_i$, consistent with previous treatments \cite{Petrov2015a,Ferrier-Barbut2016a} (also see \cite{Hu2020a}).

 \begin{figure}[htbp!] 
   \centering
   \includegraphics[width=3.4in]{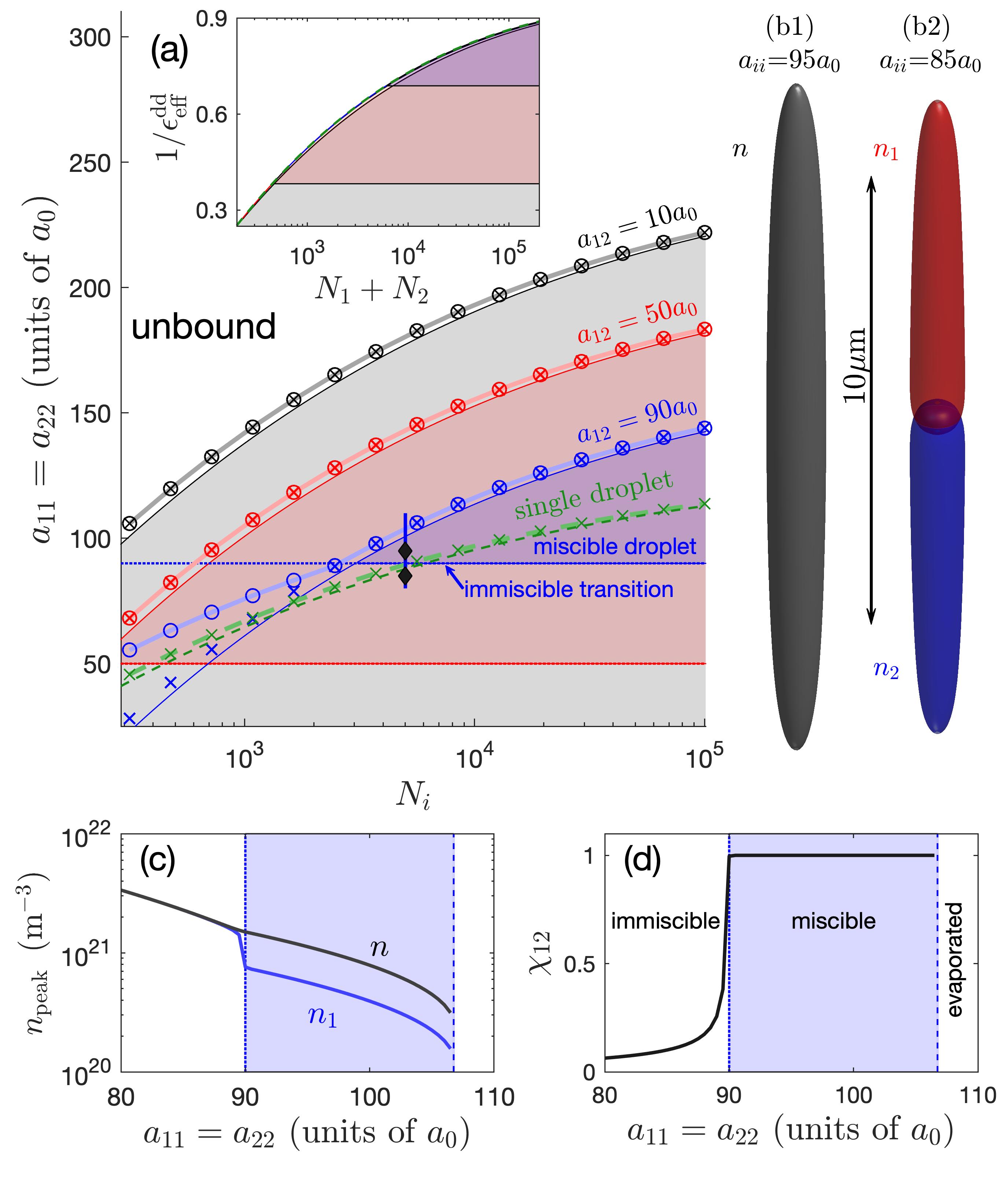}   
  \caption{Free-space droplets and miscibility in a balanced BMG. (a) Phase diagram as a function of atom number and intra-species interactions for various   $a_{12}$. BMG droplet binding predicted by variational theory (solid lines), eGPE (circles and thicker lines) and   SSA eGPE (crosses).    The immiscibility transition is at $a_{ii}=a_{12}$ (dotted horizontal lines) with  shaded regions indicating where miscible self-bound droplets are predicted by variational theory. The self-binding  for a single-component system  (green dashed lines and markers). Inset shows the variational results plotted as $1/\epsilon_\text{eff}^\text{dd}$ versus $N_1+N_2$ (the single-droplet results  as $a_{ii}/a_{ii}^\text{dd}$ versus $N_i$).
	 eGPE droplet solutions  in the (b1) miscible and (b2) immiscible regimes shown with a density isosurface at $2.5\times10^{20}\mathrm{m}^{-3}$   for  (b1)  $n=n_1+n_2$  (grey surface)  and (b2)   $n_1$ (red surface) and $n_2$ (blue surface). (c) The maximum of $n_1$ (blue line) and $n$ (black line), and  (d) the component overlap of the eGPE solution as $a_{ii}$ is varied.  The parameters for (b1), (b2) and the results in (c) and (d) are indicated in (a) by the diamonds and vertical line, respectively.  Other parameters  $a^\text{dd}_{ii}=130.8a_0$, $m=164u$, and in  (b)-(d)   $N_i=5\times10^3$.  
   \label{fig:PD}}
\end{figure}

\noindent\textbf{Quantum droplets in the balanced BMG} -
First we consider a balanced mixture with equal intra-species interactions and particle numbers in each component (i.e.~$a_{11}=a_{22}$,  $a_{11}^\text{dd}=a_{22}^\text{dd}$,  and  $N_1=N_2$). 
In the regime where both components are miscible and take the same shape (i.e.~$\psi_1=\psi_2$)  the eGPE can be reduced to an effective single component problem which we refer to as the same shape approximation (SSA):
\begin{align}
\mathcal{L}_i^\text{SSA}=h_\text{sp}+  g_\text{eff}n_i+ g_\text{eff}^\text{dd}\Phi_i+\gamma_\text{QF}n_i^{3/2},\label{LGPE1}
\end{align}
where we have set $g_\text{eff}\equiv g_{ii}+g_{12}$, and $g_\text{eff}^\text{dd}\equiv2g_{ii}^\text{dd}$ as the effective two-body contact and dipolar coupling constants, respectively. Here the quantum fluctuations ($\Delta\mu_i$) depend on the density as $n_i^{3/2}$ with coefficient \begin{align}
 \gamma_\text{QF}=\frac{2m^{3/2}}{3\pi^2\hbar^3}\left[(g_{ii}-g_{12})^{5/2}+( g^\text{dd}_\text{eff})^{5/2}\mathcal{M}_5(1/\epsilon^\text{dd}_\text{eff})\right],
 \end{align}
 where $\epsilon^\text{dd}_\text{eff}\equiv g^\text{dd}_\text{eff}/g_\text{eff}$ and $\mathcal{M}_5(x)\equiv\text{Re}\int_0^1du\,(x+3u^2-1)^{5/2}$. %
The function $\mathcal{M}_5$ can be analytically evaluated\footnote{$\mathcal{M}_5(x)=\frac{9+4x+11x^2 }{16} \sqrt{2+x} -\frac{5(1-x)^3}{32\sqrt{3}}  \Bigl[2 \log \Bigl(\sqrt{2+x}+\sqrt{3}\Bigr)-\ln | 1-x| \Bigr]$ for $x\ge-2$ and $\mathcal{M}_5(x)=0$ otherwise.} and 
has the following limits:
\begin{align}
	\begin{cases}
        (g_\text{eff}^\text{dd})^{5/2}\mathcal{M}_5(1/\epsilon^\text{dd}_\text{eff}) \to (g_\text{eff} )^{5/2},  & g_\text{eff}^\text{dd}\to 0, \\
        (g_\text{eff}^\text{dd})^{5/2}\mathcal{M}_5(1/\epsilon^\text{dd}_\text{eff}) =  (g_\text{eff} )^{5/2} \mathcal{Q}_5(\epsilon^\text{dd}_\text{eff}), & g_\text{eff} > 0,
	\end{cases}
\end{align} 
showing this expression reduces to known results for the binary contact \cite{Petrov2015a} and single component dipolar \cite{Lima2011a} condensates, respectively (see  \cite{Lima2011a}  for the definition of $\mathcal{Q}_5$ - note $\mathcal{M}_5$ is  useful because it is well-defined for $\epsilon^\text{dd}_\text{eff}\le0$).

Within the SSA, a useful  variational description is furnished by making a Gaussian ansatz for the condensate wavefunction 
$\psi_i^\text{var}=\sqrt{ {8N_i}/{\pi^{3/2}\sigma_\rho^2\sigma_z}}e^{-2(\rho^2/\sigma_\rho^2+z^2/\sigma_z^2)}$,  where $\sigma_\rho$ and $\sigma_z$ are the variational width parameters. Variational solutions are found by minimising the variational energy\footnote{Obtained from the energy functional associated with operator  in Eq.~(\ref{LGPE1}), and has similar form to the single component dipolar result of Ref.~\cite{Bisset2016a}.}
\begin{align}
\frac{E_i}{N_i}=&\frac{\hbar^2}{m}\left(\frac{2}{\sigma_{\rho}^{2}}+\frac{1}{\sigma_{z}^{2}}\right)
+\frac{4N_{i}\left[g_\text{eff}-g_\text{eff}^\text{dd}f\left(\frac{\sigma_{\rho}}{\sigma_{z}}\right)\right]}{(2\pi)^{3/2}\sigma_{\rho}^{2}\sigma_{z}}\\
&+\frac{2\gamma_\text{QF}}{5}\left(\frac{16N_{i}}{5\pi^{3/2}\sigma_{\rho}^{2}\sigma_{z}}\right)^{\frac{3}{2}},\nonumber
\end{align} 
 where  $
 f(x)=\frac{1+2x^2}{1-x^2}-\frac{3x^2\tanh^{-1}\sqrt{1-x^2}}{(1-x^2)^{3/2}}$. 
 
In free-space (i.e.~$V=0$) self-bound droplets are stable ground states where their energy is negative, otherwise they are unstable to evaporating to the trivial solution  ($\psi_i\to0$).
 In Fig.~\ref{fig:PD}(a) we present a phase diagram obtained from the variational theory with $E_i=0$ taken to define the stable droplet boundary.  This shows that, for a given atom number, a stable droplet will occur when the intra-species scattering length is below some threshold value. This threshold tends to decrease with increasing $a_{12}$, as the inter-species repulsion competes against the attractive DDIs. 
 For reference the self-binding boundary for a single component droplet\footnote{I.e.~for a single component system with parameters $N_i$, $a_{ii}$ and   $a_{ii}^\text{dd}$.}  \cite{Baillie2016b} is also indicated.  This shows that the BMG droplets can be stable in regimes where each component could not form a stable droplet by itself. In the inset we present the phase diagram rescaled to the effective parameters and show that the boundaries approximately collapse to a single curve. 
 
In Fig.~\ref{fig:PD}(b1) we show an example of a miscible droplet obtained by solving the eGPE, in a regime where both components have identical wavefunctions. In this case the full eGPE is equivalent to the SSA eGPE [see Eqs.~(\ref{LGPEi}) and (\ref{LGPE1})]. 
 As the value of $a_{12}$ is lowered, the two components become immiscible and phase-separate -- thus the SSA is no longer applicable. The two separated components orient in a head-to-tail configuration to minimise the DDI energy  [see Fig.~\ref{fig:PD}(b2)].

For the balanced system in the miscible regime (where the SSA applies) the DDIs are identical within and between the components. Thus immiscibility is determined by the short-ranged interactions with the transition occurring when $a_{12}=\sqrt{a_{11}a_{22}}$ \cite{Mineev1974a}. This allows us determine the regions (shaded) of the phase diagram where the BMG droplets are miscible.  In  Fig.~\ref{fig:PD}(c) and (d) we show the peak densities and the overlap $\chi_{12}\equiv\frac{1}{\sqrt{N_1N_2}}\int d\bx\,\psi_1^*\psi_2$ (quantifying miscibility) as the transition is crossed by varying $a_{ii}$  for $a_{12}=90a_0$.
 
 We have calculated stationary droplet states of the full eGPE and the SSA eGPE to determine the self-binding phase boundary.  
 These results are generally in good agreement with, although lie slightly above, the variational  boundaries.  An exception is for  $a_{12}=90a_0$ and $N_1\lesssim3\times10^3$. In this region the variational boundary is below the immiscibility line, and the droplets are immiscible at the boundary. As such both the variational and SSA eGPE are inapplicable, and the full eGPE predicts a significantly higher phase boundary. Here the self-binding line is similar to the single droplet case --   expected since the droplets components have spatially separated -- but it is shifted upwards (relative to the single-droplet boundary) due to the stabilizing attractive DDI  between the components.

\begin{figure}[tbp] 
   \centering
   \includegraphics[width=3.2in]{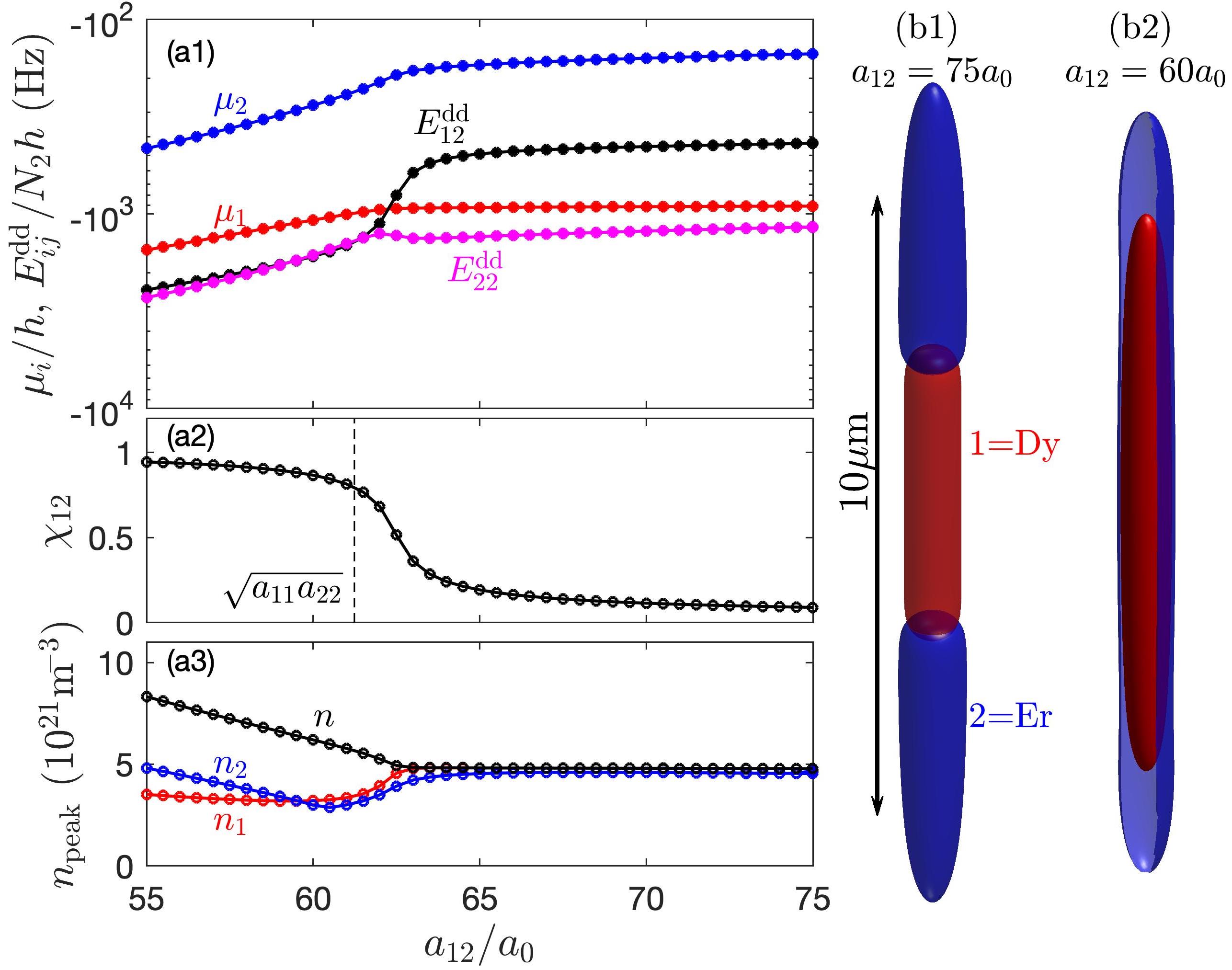}   
  \caption{Immiscibility transition of a free-space BMG droplet comprised of $N_1=5\times10^3$ Dy atoms and  $N_2= 10^4$ Er atoms.  (a1)-(a3) Chemical potentials, DDI energies, component overlap and peak densities as the interspecies scattering length is changed across the immiscibility transition.  (b1), (b2) Density isosurfaces at $5\times10^{20}/$m$^3$ of the ground states at labeled $a_{12}$ values [component 2 is cut away in (b2) to reveal component 1]. 
	Other parameters:   $\{a_{11}, a_{22},a^\text{dd}_{11},a^\text{dd}_{22}\}= \{75, 50, 130.8,65.5\}a_0$, and  $m=165u$.  
   \label{fig:imbalanced}}
\end{figure}

\begin{figure}[tbp] 
   \centering
   \includegraphics[width=3.2in]{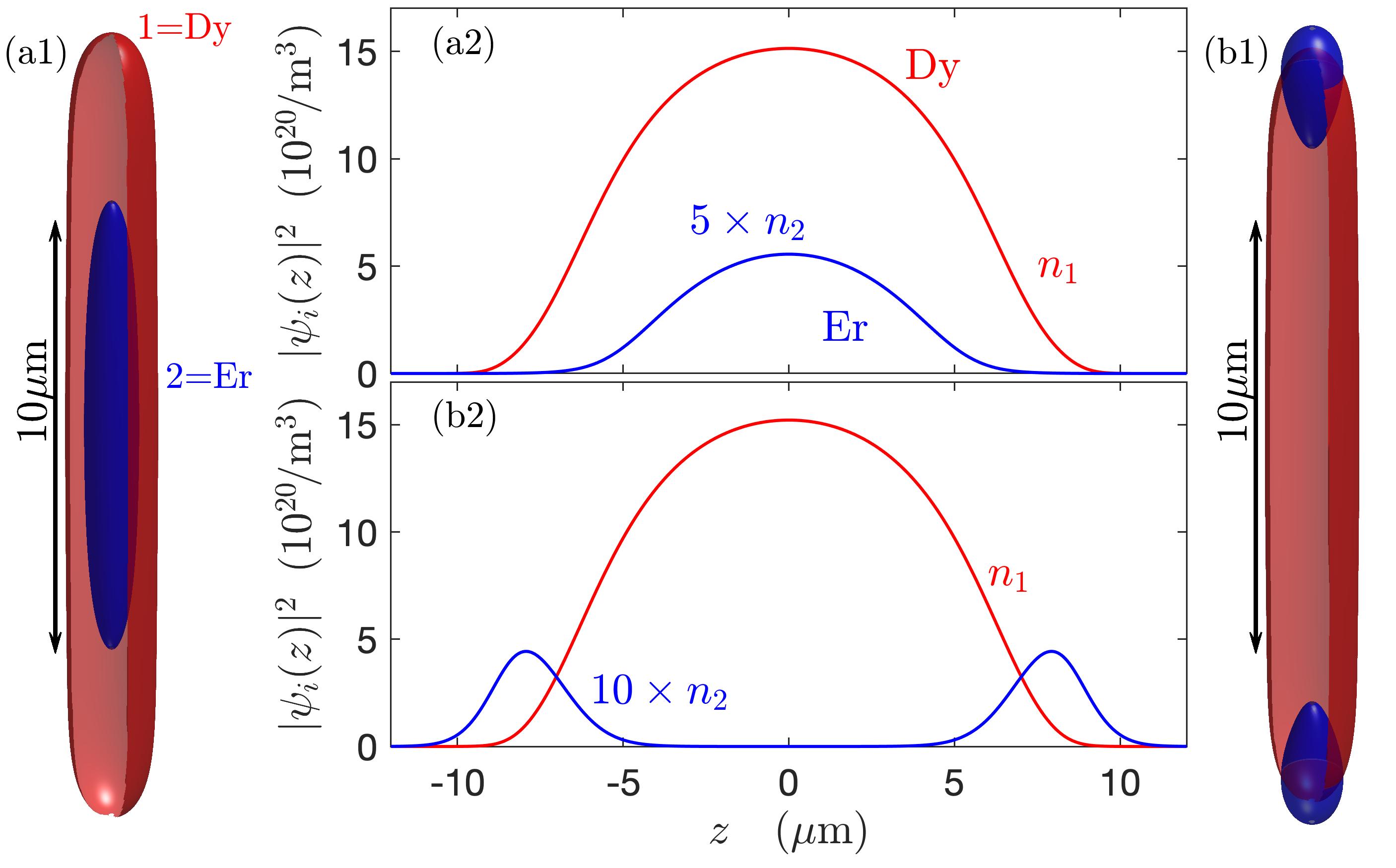} 
  \caption{Er condensate confined by a Dy droplet in free-space. (a1,b1) Dy  (red) and  Er  (blue) density isosurfaces  at $2\times10^{19}$m$^{-3}$, and (a2,b2) the respective density on the $z$-axis. In (a)  $a_{12}=65a_0$ and in (b) $a_{12}=70a_0$. 
	Other parameters: $N_1=10^4$, $N_2=500$,  $\{a_{11},a_{22},a^\text{dd}_{11},a^\text{dd}_{22}\}=\{90,80,130.8,65.5\}a_0$,  and  $m=165u$.  
   \label{fig:qtt}}
\end{figure}

\noindent\textbf{Quantum droplets in an imbalanced BMG} -
Our formalism applies to the general imbalanced case where the atom number and intra-species interactions are different. In Fig.~\ref{fig:imbalanced}  we consider the ground state droplet properties of Dy-Er BMG as the inter-species scattering length is ramped through the immiscibility transition.  These solutions are obtained by numerically solving for the ground states of the full eGPE.

The chemical potential $\mu_i$ is the energy required to add a particle of component-$i$  and provides a useful characterization of the system. The values of $\mu_i$ in Fig.~\ref{fig:imbalanced}(a1) are negative, indicating that both components are self-bound. In the regime we examine, $\mu_1$ (for Dy)   is less than $\mu_2$ (for  Er).   We also see that $\mu_i$ increases with increasing $a_{12}$  when the system is miscible [i.e.~where $\chi_{12}\sim1$, see (a2)], since $a_{12}$ makes an important contribution to the energy when the components  spatially overlap. When the system transitions to being immiscible, the chemical potentials become almost independent of $a_{12}$. However, the DDIs between the components still play an important role. To quantify this we show the inter- and intra-species DDI energies  $E^\text{dd}_{ij}=\frac{1}{2}g_{ij}^\text{dd}\int d\bx\,n_i\Phi_j$ in Fig.~\ref{fig:imbalanced}(a1).  We see that, even when the system is  immiscible,  $|E^\text{dd}_{12}/N_2|>|\mu_2|$, i.e.~the inter-species DDI energy per particle is larger (in magnitude) than the Er chemical potential. This emphasises the often dominant role of long-ranged inter-component interactions in this system.

The Dy density is slightly higher than the Er density in the immiscible regime  [$a_{12}\gtrsim60a_0$ in Fig.~\ref{fig:imbalanced}(a3)], and the phase-separated droplet organizes to have the Dy component at the center with Er component above and below [see Fig.~\ref{fig:imbalanced}(b1)].   By increasing $a_{11}$ we find similar behavior to Fig.~\ref{fig:imbalanced}, but the Dy density decreases (in the immiscible regime) and the Er component moves to the center of the droplet.

In Fig.~\ref{fig:qtt} we consider a more strongly imbalanced case of $10^4$ Dy atoms and  500 Er atoms. Taking  $a_{11}=90a_0$ ($<a_{11}^\text{dd}$) for Dy, it is in a regime where it forms a self-bound droplet by itself. For  Er  $a_{22}=80a_0$ ($>a_{22}^\text{dd}$) so that its overall intra-species interactions are contact dominated and repulsive (i.e.~it will not bind into a droplet) and can be considered as a regular dipolar condensate that would evaporate in free-space.
For the combined system, the inter-species interactions allow the Dy droplet to bind the Er atoms. For the case in Fig.~\ref{fig:qtt}(a1,a2)  the components are miscible and the Er atoms sit in the middle of the Dy droplet,  while the case in Fig.~\ref{fig:qtt}(b1,b2) (with higher $a_{12}$) is immiscible, and the Er atoms are instead confined to the effective potential minima at the top and bottom of the Dy droplet arising from the attractive DDIs. In both cases the Dy component is only weakly affected by the Er component which is more than an order of magnitude less dense.

\begin{figure}[tbp] 
   \centering
   \includegraphics[trim=140 0 0 0,clip=true,width=3.5in]{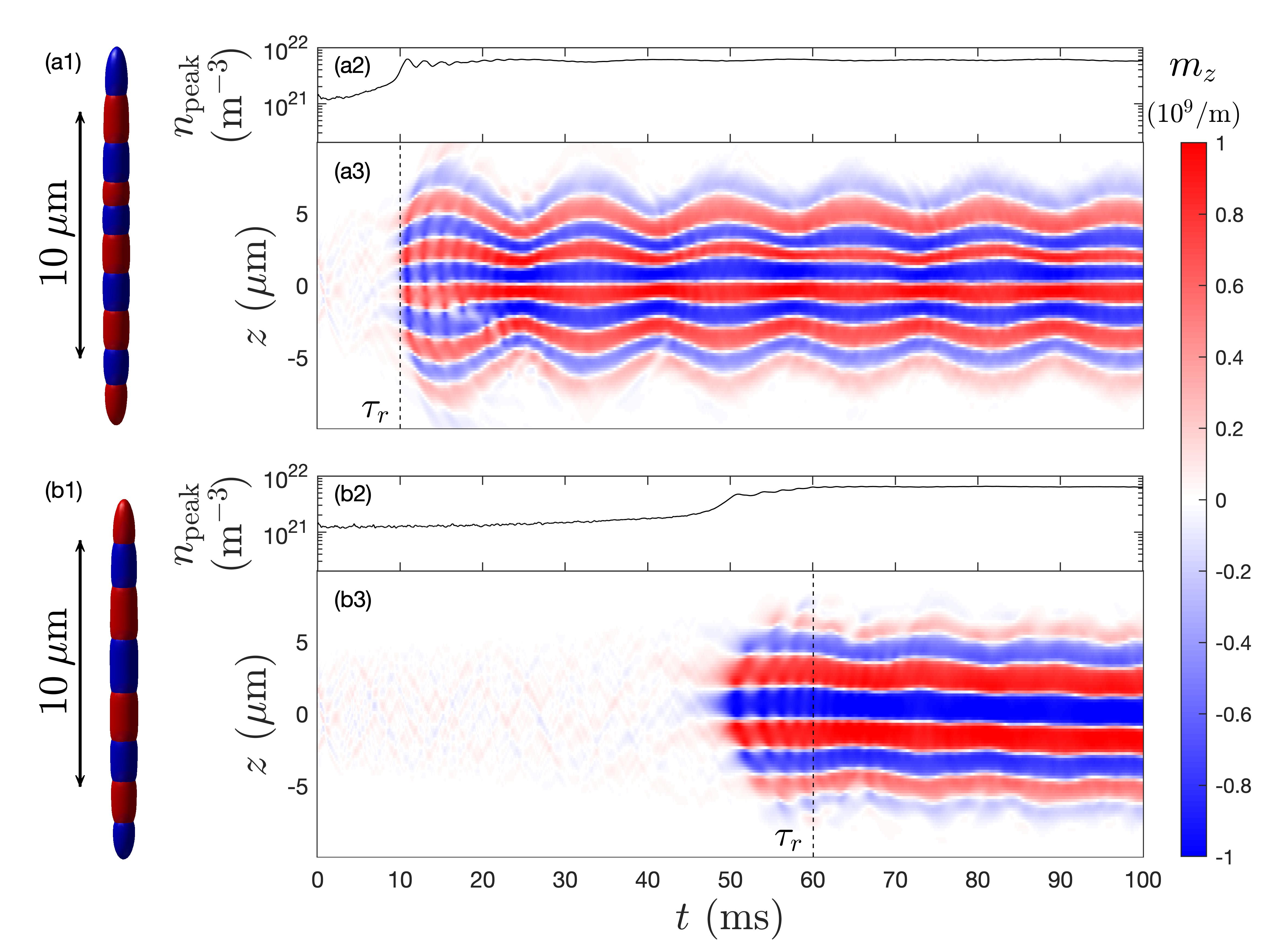} 
	\caption{Preparation of BMG droplets from an initial miscible Dy-Dy condensate. (a1)-(a3) Results from a $\tau_r=10\,$ms ramp. (a1) $10^{20}/$m$^3$ density isosurfaces of component 1 (red) and 2 (blue) at $t=100\,$ms.  Evolution of (a2) peak total density  $n_\mathrm{peak}$ and (a3) pseudo-spin density $m_z$. (b1)-(b3) Similar analysis for a $\tau_r=60\,$ms ramp. Initial and final values for the intra-species scattering lengths are  $a_{ii}=110a_0$ and  $a_{ii}^\text{f}=80a_0$, respectively. The initial harmonic confinement is $V=\tfrac{1}{2}m\sum_{\nu}\omega_\nu^2x_\nu^2$, with $\omega_{x,y}/2\pi=250$Hz, and $\omega_z/2\pi=125$Hz.   
  Other parameters: $N_{i}=5\times10^3$,   $a^\text{dd}_{ii}=130.8a_0$,  $a_{12}=90a_0$  and  $m=163.9u$.  
   \label{fig:dynamics}}
\end{figure}

\noindent\textbf{Preparation dynamics of BMG droplets} - Here we simulate a procedure to prepare a BMG droplet, starting from a trapped condensate in the miscible regime, and then reducing the scattering lengths into the droplet regime while removing the trap potential (cf.~\cite{Chomaz2016a,Schmitt2016a}). 
We use the time-dependent eGPE $i\hbar\dot\psi_i=\mathcal{L}_i\psi_i$ from an initial state corresponding to a harmonically trapped condensate with noise added to mimic vacuum fluctuations \cite{cfieldRev2008}, following the procedure described in Ref.~\cite{Bisset2015a} (also see \cite{Chomaz2019a,Natale2019a}). At $t=0$ we ramp into the droplet regime over a time period of $\tau_r$, by linearly reducing the intra-species scattering lengths $a_{ii}$ to a final value  $a_{ii}^\text{f}$, and linearly reducing the trap frequencies to zero. These parameters are then held constant for the remainder of the simulation.

For $ a_{ii}^\text{f}$ in the range $(90a_0,103a_0)$  the free-space ground state is a miscible BMG droplet (cf.~Fig.~\ref{fig:PD}). Simulations of this case (not shown) successfully generate a self-bound droplet with collective oscillations excited by the finite ramp time (similar to the single component simulations in Ref.~\cite{Baillie2016b}). The droplet produced lasts for the duration of our simulations  ($\gtrsim100\,$ms) without any noticeable decay. The inclusion of three-body loss would limit the droplet lifetime.

For $a_{ii}^\text{f}\lesssim90a_0$ the  ground state  droplet is immiscible with two domains [cf.~Fig.~\ref{fig:PD}(b2)]. We show results simulating the evolution in  this regime for $\tau_r=10\,$ms and $60\,$ms  in Fig.~\ref{fig:dynamics}. The droplet  produced by the ramp  has a greater number of domains, with a nearly periodic spatial pattern. We characterize the domain formation and its evolution with the (pseudo)-spin density $m_z(z)\equiv\int dx\,dy\,(|\psi_{1}|^2-|\psi_2|^2$) in Fig.~\ref{fig:dynamics}(a3) and (b3). This observable reveals the domains appearing near the conclusion of the ramp, and the breathing-mode-like collective excitation of these domains along the $z$ axis. Note that we dot no observe any merging or decay of the domains once they form even under these strong collective dynamics, demonstrating that these structures are robust. For the slower ramp we find that fewer domains are produced. This $\tau_r$-dependence  is similar to that found in studies of the domain formation in the immiscibility transition of a binary condensate
\cite{Sabbatini2011a,Nicklas2015a}, which was described using the Kibble-Zurek mechanism. This suggests that the additional domains we observe are defects arising from non-adiabatically crossing the phase transition.

\noindent\textbf{Conclusions and outlook} -
We have developed a theory for a new category of quantum droplet that exhibits the rich behavior of binary fluids.  Using this theory we have characterized a droplet phase diagram and the properties of miscible and immiscible droplets. We have shown that a single magnetic droplet can be used to bind a magnetic condensate, and thus potentially even individual magnetic atoms. This test tube-like behavior has similarities to helium nanodroplets,  which act as quantum solvents to attach other chemical species, e.g.~for mK spectroscopy \cite{Stienkemeier2006a}.
We have also considered the dynamics of forming a BMG droplet from an initial trapped condensate in the miscible regime, showing that an interaction and trap ramp can be used to prepare self-bound droplets on the typical time scales available to experiments. Crossing the immiscibility transition at a finite rate leads to additional  domains being created compared to the ground state case. Pathways to produce a droplet closer to the ground state case include (i) using a slower ramp, (ii) ramping down from an immiscible condensate (i.e.~already having several domains), or (iii) directly evaporatively cooling into the droplet regime (cf.~Ref.~\cite{Chomaz2019a}).

\noindent\textbf{Acknowledgments} -We acknowledge support from the Marsden Fund of the Royal Society of New Zealand and useful discussions with Au-Chen Lee.


\begin{thebibliography}{22}%
\makeatletter
\providecommand \@ifxundefined [1]{%
 \@ifx{#1\undefined}
}%
\providecommand \@ifnum [1]{%
 \ifnum #1\expandafter \@firstoftwo
 \else \expandafter \@secondoftwo
 \fi
}%
\providecommand \@ifx [1]{%
 \ifx #1\expandafter \@firstoftwo
 \else \expandafter \@secondoftwo
 \fi
}%
\providecommand \natexlab [1]{#1}%
\providecommand \enquote  [1]{``#1''}%
\providecommand \bibnamefont  [1]{#1}%
\providecommand \bibfnamefont [1]{#1}%
\providecommand \citenamefont [1]{#1}%
\providecommand \href@noop [0]{\@secondoftwo}%
\providecommand \href [0]{\begingroup \@sanitize@url \@href}%
\providecommand \@href[1]{\@@startlink{#1}\@@href}%
\providecommand \@@href[1]{\endgroup#1\@@endlink}%
\providecommand \@sanitize@url [0]{\catcode `\\12\catcode `\$12\catcode
  `\&12\catcode `\#12\catcode `\^12\catcode `\_12\catcode `\%12\relax}%
\providecommand \@@startlink[1]{}%
\providecommand \@@endlink[0]{}%
\providecommand \url  [0]{\begingroup\@sanitize@url \@url }%
\providecommand \@url [1]{\endgroup\@href {#1}{\urlprefix }}%
\providecommand \urlprefix  [0]{URL }%
\providecommand \Eprint [0]{\href }%
\providecommand \doibase [0]{http://dx.doi.org/}%
\providecommand \selectlanguage [0]{\@gobble}%
\providecommand \bibinfo  [0]{\@secondoftwo}%
\providecommand \bibfield  [0]{\@secondoftwo}%
\providecommand \translation [1]{[#1]}%
\providecommand \BibitemOpen [0]{}%
\providecommand \bibitemStop [0]{}%
\providecommand \bibitemNoStop [0]{.\EOS\space}%
\providecommand \EOS [0]{\spacefactor3000\relax}%
\providecommand \BibitemShut  [1]{\csname bibitem#1\endcsname}%
\let\auto@bib@innerbib\@empty
\bibitem [{\citenamefont {Bulgac}(2002)}]{Bulgac2002a}%
  \BibitemOpen
  \bibfield  {author} {\bibinfo {author} {\bibfnamefont {Aurel}\ \bibnamefont
  {Bulgac}},\ }\bibfield  {title} {\enquote {\bibinfo {title} {Dilute quantum
  droplets},}\ }\href {\doibase 10.1103/PhysRevLett.89.050402} {\bibfield
  {journal} {\bibinfo  {journal} {Phys. Rev. Lett.}\ }\textbf {\bibinfo
  {volume} {89}},\ \bibinfo {pages} {050402} (\bibinfo {year}
  {2002})}\BibitemShut {NoStop}%
\bibitem [{\citenamefont {Petrov}(2015)}]{Petrov2015a}%
  \BibitemOpen
  \bibfield  {author} {\bibinfo {author} {\bibfnamefont {D.~S.}\ \bibnamefont
  {Petrov}},\ }\bibfield  {title} {\enquote {\bibinfo {title} {Quantum
  mechanical stabilization of a collapsing {B}ose-{B}ose mixture},}\ }\href
  {\doibase 10.1103/PhysRevLett.115.155302} {\bibfield  {journal} {\bibinfo
  {journal} {Phys. Rev. Lett.}\ }\textbf {\bibinfo {volume} {115}},\ \bibinfo
  {pages} {155302} (\bibinfo {year} {2015})}\BibitemShut {NoStop}%
\bibitem [{\citenamefont {Cabrera}\ \emph {et~al.}(2018)\citenamefont
  {Cabrera}, \citenamefont {Tanzi}, \citenamefont {Sanz}, \citenamefont
  {Naylor}, \citenamefont {Thomas}, \citenamefont {Cheiney},\ and\
  \citenamefont {Tarruell}}]{Cabrera2018a}%
  \BibitemOpen
  \bibfield  {author} {\bibinfo {author} {\bibfnamefont {C.~R.}\ \bibnamefont
  {Cabrera}}, \bibinfo {author} {\bibfnamefont {L.}~\bibnamefont {Tanzi}},
  \bibinfo {author} {\bibfnamefont {J.}~\bibnamefont {Sanz}}, \bibinfo {author}
  {\bibfnamefont {B.}~\bibnamefont {Naylor}}, \bibinfo {author} {\bibfnamefont
  {P.}~\bibnamefont {Thomas}}, \bibinfo {author} {\bibfnamefont
  {P.}~\bibnamefont {Cheiney}}, \ and\ \bibinfo {author} {\bibfnamefont
  {L.}~\bibnamefont {Tarruell}},\ }\bibfield  {title} {\enquote {\bibinfo
  {title} {Quantum liquid droplets in a mixture of {B}ose-{E}instein
  condensates},}\ }\href {\doibase 10.1126/science.aao5686} {\bibfield
  {journal} {\bibinfo  {journal} {Science}\ }\textbf {\bibinfo {volume}
  {359}},\ \bibinfo {pages} {301--304} (\bibinfo {year} {2018})}\BibitemShut
  {NoStop}%
\bibitem [{\citenamefont {Semeghini}\ \emph {et~al.}(2018)\citenamefont
  {Semeghini}, \citenamefont {Ferioli}, \citenamefont {Masi}, \citenamefont
  {Mazzinghi}, \citenamefont {Wolswijk}, \citenamefont {Minardi}, \citenamefont
  {Modugno}, \citenamefont {Modugno}, \citenamefont {Inguscio},\ and\
  \citenamefont {Fattori}}]{Semeghini2018a}%
  \BibitemOpen
  \bibfield  {author} {\bibinfo {author} {\bibfnamefont {G.}~\bibnamefont
  {Semeghini}}, \bibinfo {author} {\bibfnamefont {G.}~\bibnamefont {Ferioli}},
  \bibinfo {author} {\bibfnamefont {L.}~\bibnamefont {Masi}}, \bibinfo {author}
  {\bibfnamefont {C.}~\bibnamefont {Mazzinghi}}, \bibinfo {author}
  {\bibfnamefont {L.}~\bibnamefont {Wolswijk}}, \bibinfo {author}
  {\bibfnamefont {F.}~\bibnamefont {Minardi}}, \bibinfo {author} {\bibfnamefont
  {M.}~\bibnamefont {Modugno}}, \bibinfo {author} {\bibfnamefont
  {G.}~\bibnamefont {Modugno}}, \bibinfo {author} {\bibfnamefont
  {M.}~\bibnamefont {Inguscio}}, \ and\ \bibinfo {author} {\bibfnamefont
  {M.}~\bibnamefont {Fattori}},\ }\bibfield  {title} {\enquote {\bibinfo
  {title} {Self-bound quantum droplets of atomic mixtures in free space},}\
  }\href {\doibase 10.1103/PhysRevLett.120.235301} {\bibfield  {journal}
  {\bibinfo  {journal} {Phys. Rev. Lett.}\ }\textbf {\bibinfo {volume} {120}},\
  \bibinfo {pages} {235301} (\bibinfo {year} {2018})}\BibitemShut {NoStop}%
\bibitem [{\citenamefont {D'Errico}\ \emph {et~al.}(2019)\citenamefont
  {D'Errico}, \citenamefont {Burchianti}, \citenamefont {Prevedelli},
  \citenamefont {Salasnich}, \citenamefont {Ancilotto}, \citenamefont
  {Modugno}, \citenamefont {Minardi},\ and\ \citenamefont
  {Fort}}]{DErrico2019a}%
  \BibitemOpen
  \bibfield  {author} {\bibinfo {author} {\bibfnamefont {C.}~\bibnamefont
  {D'Errico}}, \bibinfo {author} {\bibfnamefont {A.}~\bibnamefont
  {Burchianti}}, \bibinfo {author} {\bibfnamefont {M.}~\bibnamefont
  {Prevedelli}}, \bibinfo {author} {\bibfnamefont {L.}~\bibnamefont
  {Salasnich}}, \bibinfo {author} {\bibfnamefont {F.}~\bibnamefont
  {Ancilotto}}, \bibinfo {author} {\bibfnamefont {M.}~\bibnamefont {Modugno}},
  \bibinfo {author} {\bibfnamefont {F.}~\bibnamefont {Minardi}}, \ and\
  \bibinfo {author} {\bibfnamefont {C.}~\bibnamefont {Fort}},\ }\bibfield
  {title} {\enquote {\bibinfo {title} {Observation of quantum droplets in a
  heteronuclear bosonic mixture},}\ }\href {\doibase %
  10.1103/PhysRevResearch.1.033155} {\bibfield  {journal} {\bibinfo  {journal}
  {Phys. Rev. Research}\ }\textbf {\bibinfo {volume} {1}},\ \bibinfo {pages}
  {033155} (\bibinfo {year} {2019})}\BibitemShut {NoStop}%
\bibitem [{\citenamefont {Ferrier-Barbut}\ \emph {et~al.}(2016)\citenamefont
  {Ferrier-Barbut}, \citenamefont {Kadau}, \citenamefont {Schmitt},
  \citenamefont {Wenzel},\ and\ \citenamefont {Pfau}}]{Ferrier-Barbut2016a}%
  \BibitemOpen
  \bibfield  {author} {\bibinfo {author} {\bibfnamefont {Igor}\ \bibnamefont
  {Ferrier-Barbut}}, \bibinfo {author} {\bibfnamefont {Holger}\ \bibnamefont
  {Kadau}}, \bibinfo {author} {\bibfnamefont {Matthias}\ \bibnamefont
  {Schmitt}}, \bibinfo {author} {\bibfnamefont {Matthias}\ \bibnamefont
  {Wenzel}}, \ and\ \bibinfo {author} {\bibfnamefont {Tilman}\ \bibnamefont
  {Pfau}},\ }\bibfield  {title} {\enquote {\bibinfo {title} {Observation of
  quantum droplets in a strongly dipolar {B}ose gas},}\ }\href {\doibase %
  10.1103/PhysRevLett.116.215301} {\bibfield  {journal} {\bibinfo  {journal}
  {Phys. Rev. Lett.}\ }\textbf {\bibinfo {volume} {116}},\ \bibinfo {pages}
  {215301} (\bibinfo {year} {2016})}\BibitemShut {NoStop}%
\bibitem [{\citenamefont {Chomaz}\ \emph {et~al.}(2016)\citenamefont {Chomaz},
  \citenamefont {Baier}, \citenamefont {Petter}, \citenamefont {Mark},
  \citenamefont {W\"achtler}, \citenamefont {Santos},\ and\ \citenamefont
  {Ferlaino}}]{Chomaz2016a}%
  \BibitemOpen
  \bibfield  {author} {\bibinfo {author} {\bibfnamefont {L.}~\bibnamefont
  {Chomaz}}, \bibinfo {author} {\bibfnamefont {S.}~\bibnamefont {Baier}},
  \bibinfo {author} {\bibfnamefont {D.}~\bibnamefont {Petter}}, \bibinfo
  {author} {\bibfnamefont {M.~J.}\ \bibnamefont {Mark}}, \bibinfo {author}
  {\bibfnamefont {F.}~\bibnamefont {W\"achtler}}, \bibinfo {author}
  {\bibfnamefont {L.}~\bibnamefont {Santos}}, \ and\ \bibinfo {author}
  {\bibfnamefont {F.}~\bibnamefont {Ferlaino}},\ }\bibfield  {title} {\enquote
  {\bibinfo {title} {Quantum-fluctuation-driven crossover from a dilute
  {B}ose-{E}instein condensate to a macrodroplet in a dipolar quantum fluid},}\
  }\href {\doibase 10.1103/PhysRevX.6.041039} {\bibfield  {journal} {\bibinfo
  {journal} {Phys. Rev. X}\ }\textbf {\bibinfo {volume} {6}},\ \bibinfo {pages}
  {041039} (\bibinfo {year} {2016})}\BibitemShut {NoStop}%
\bibitem [{\citenamefont {Trautmann}\ \emph {et~al.}(2018)\citenamefont
  {Trautmann}, \citenamefont {Ilzh\"ofer}, \citenamefont {Durastante},
  \citenamefont {Politi}, \citenamefont {Sohmen}, \citenamefont {Mark},\ and\
  \citenamefont {Ferlaino}}]{Trautmann2018a}%
  \BibitemOpen
  \bibfield  {author} {\bibinfo {author} {\bibfnamefont {A.}~\bibnamefont
  {Trautmann}}, \bibinfo {author} {\bibfnamefont {P.}~\bibnamefont
  {Ilzh\"ofer}}, \bibinfo {author} {\bibfnamefont {G.}~\bibnamefont
  {Durastante}}, \bibinfo {author} {\bibfnamefont {C.}~\bibnamefont {Politi}},
  \bibinfo {author} {\bibfnamefont {M.}~\bibnamefont {Sohmen}}, \bibinfo
  {author} {\bibfnamefont {M.~J.}\ \bibnamefont {Mark}}, \ and\ \bibinfo
  {author} {\bibfnamefont {F.}~\bibnamefont {Ferlaino}},\ }\bibfield  {title}
  {\enquote {\bibinfo {title} {Dipolar quantum mixtures of erbium and
  dysprosium atoms},}\ }\href {\doibase 10.1103/PhysRevLett.121.213601}
  {\bibfield  {journal} {\bibinfo  {journal} {Phys. Rev. Lett.}\ }\textbf
  {\bibinfo {volume} {121}},\ \bibinfo {pages} {213601} (\bibinfo {year}
  {2018})}\BibitemShut {NoStop}%
\bibitem [{\citenamefont {{Hu}}\ and\ \citenamefont {{Liu}}()}]{Hu2020a}%
  \BibitemOpen
  \bibfield  {author} {\bibinfo {author} {\bibfnamefont {Hui}\ \bibnamefont
  {{Hu}}}\ and\ \bibinfo {author} {\bibfnamefont {Xia-Ji}\ \bibnamefont
  {{Liu}}},\ }\bibfield  {title} {\enquote {\bibinfo {title} {{Consistent
  theory of self-bound quantum droplets with bosonic pairing}},}\ }\href@noop
  {} {\ }\Eprint {http://arxiv.org/abs/2005.08581} {arXiv:2005.08581}
  \BibitemShut {NoStop}%
\bibitem [{\citenamefont {Lima}\ and\ \citenamefont
  {Pelster}(2011)}]{Lima2011a}%
  \BibitemOpen
  \bibfield  {author} {\bibinfo {author} {\bibfnamefont {Aristeu R.~P.}\
  \bibnamefont {Lima}}\ and\ \bibinfo {author} {\bibfnamefont {Axel}\
  \bibnamefont {Pelster}},\ }\bibfield  {title} {\enquote {\bibinfo {title}
  {Quantum fluctuations in dipolar {B}ose gases},}\ }\href {\doibase %
  10.1103/PhysRevA.84.041604} {\bibfield  {journal} {\bibinfo  {journal} {Phys.
  Rev. A}\ }\textbf {\bibinfo {volume} {84}},\ \bibinfo {pages} {041604}
  (\bibinfo {year} {2011})}\BibitemShut {NoStop}%
\bibitem [{\citenamefont {Bisset}\ \emph {et~al.}(2016)\citenamefont {Bisset},
  \citenamefont {Wilson}, \citenamefont {Baillie},\ and\ \citenamefont
  {Blakie}}]{Bisset2016a}%
  \BibitemOpen
  \bibfield  {author} {\bibinfo {author} {\bibfnamefont {R.~N.}\ \bibnamefont
  {Bisset}}, \bibinfo {author} {\bibfnamefont {R.~M.}\ \bibnamefont {Wilson}},
  \bibinfo {author} {\bibfnamefont {D.}~\bibnamefont {Baillie}}, \ and\
  \bibinfo {author} {\bibfnamefont {P.~B.}\ \bibnamefont {Blakie}},\ }\bibfield
   {title} {\enquote {\bibinfo {title} {Ground-state phase diagram of a dipolar
  condensate with quantum fluctuations},}\ }\href {\doibase %
  10.1103/PhysRevA.94.033619} {\bibfield  {journal} {\bibinfo  {journal} {Phys.
  Rev. A}\ }\textbf {\bibinfo {volume} {94}},\ \bibinfo {pages} {033619}
  (\bibinfo {year} {2016})}\BibitemShut {NoStop}%
\bibitem [{\citenamefont {Baillie}\ \emph {et~al.}(2016)\citenamefont
  {Baillie}, \citenamefont {Wilson}, \citenamefont {Bisset},\ and\
  \citenamefont {Blakie}}]{Baillie2016b}%
  \BibitemOpen
  \bibfield  {author} {\bibinfo {author} {\bibfnamefont {D.}~\bibnamefont
  {Baillie}}, \bibinfo {author} {\bibfnamefont {R.~M.}\ \bibnamefont {Wilson}},
  \bibinfo {author} {\bibfnamefont {R.~N.}\ \bibnamefont {Bisset}}, \ and\
  \bibinfo {author} {\bibfnamefont {P.~B.}\ \bibnamefont {Blakie}},\ }\bibfield
   {title} {\enquote {\bibinfo {title} {Self-bound dipolar droplet: A localized
  matter wave in free space},}\ }\href {\doibase 10.1103/PhysRevA.94.021602}
  {\bibfield  {journal} {\bibinfo  {journal} {Phys. Rev. A}\ }\textbf {\bibinfo
  {volume} {94}},\ \bibinfo {pages} {021602(R)} (\bibinfo {year}
  {2016})}\BibitemShut {NoStop}%
\bibitem [{\citenamefont {Mineev}(1974)}]{Mineev1974a}%
  \BibitemOpen
  \bibfield  {author} {\bibinfo {author} {\bibfnamefont {V.P.}\ \bibnamefont
  {Mineev}},\ }\bibfield  {title} {\enquote {\bibinfo {title} {The theory of
  the solution of two near-ideal {Bose} gases},}\ }\href
  {http://jetp.ac.ru/cgi-bin/e/index/r/67/1/p263?a=list} {\bibfield  {journal}
  {\bibinfo  {journal} {Zh. Eksp. Teor. Fiz.}\ }\textbf {\bibinfo {volume}
  {67}},\ \bibinfo {pages} {263} (\bibinfo {year} {1974})}\BibitemShut
  {NoStop}%
\bibitem [{\citenamefont {Schmitt}\ \emph {et~al.}(2016)\citenamefont
  {Schmitt}, \citenamefont {Wenzel}, \citenamefont {B{\"o}ttcher},
  \citenamefont {Ferrier-Barbut},\ and\ \citenamefont {Pfau}}]{Schmitt2016a}%
  \BibitemOpen
  \bibfield  {author} {\bibinfo {author} {\bibfnamefont {Matthias}\
  \bibnamefont {Schmitt}}, \bibinfo {author} {\bibfnamefont {Matthias}\
  \bibnamefont {Wenzel}}, \bibinfo {author} {\bibfnamefont {Fabian}\
  \bibnamefont {B{\"o}ttcher}}, \bibinfo {author} {\bibfnamefont {Igor}\
  \bibnamefont {Ferrier-Barbut}}, \ and\ \bibinfo {author} {\bibfnamefont
  {Tilman}\ \bibnamefont {Pfau}},\ }\bibfield  {title} {\enquote {\bibinfo
  {title} {Self-bound droplets of a dilute magnetic quantum liquid},}\ }\href
  {http://dx.doi.org/10.1038/nature20126} {\bibfield  {journal} {\bibinfo
  {journal} {Nature}\ }\textbf {\bibinfo {volume} {539}},\ \bibinfo {pages}
  {259--262} (\bibinfo {year} {2016})}\BibitemShut {NoStop}%
\bibitem [{\citenamefont {Blakie}\ \emph {et~al.}(2008)\citenamefont {Blakie},
  \citenamefont {Bradley}, \citenamefont {Davis}, \citenamefont {Ballagh},\
  and\ \citenamefont {Gardiner}}]{cfieldRev2008}%
  \BibitemOpen
  \bibfield  {author} {\bibinfo {author} {\bibfnamefont {P.~B.}\ \bibnamefont
  {Blakie}}, \bibinfo {author} {\bibfnamefont {A.~S.}\ \bibnamefont {Bradley}},
  \bibinfo {author} {\bibfnamefont {M.~J.}\ \bibnamefont {Davis}}, \bibinfo
  {author} {\bibfnamefont {R.~J.}\ \bibnamefont {Ballagh}}, \ and\ \bibinfo
  {author} {\bibfnamefont {C.~W.}\ \bibnamefont {Gardiner}},\ }\bibfield
  {title} {\enquote {\bibinfo {title} {Dynamics and statistical mechanics of
  ultra-cold {B}ose gases using c-field techniques},}\ }\href {\doibase %
  10.1080/00018730802564254} {\bibfield  {journal} {\bibinfo  {journal} {Adv.
  Phys.}\ }\textbf {\bibinfo {volume} {57}},\ \bibinfo {pages} {363} (\bibinfo
  {year} {2008})}\BibitemShut {NoStop}%
\bibitem [{\citenamefont {Bisset}\ and\ \citenamefont
  {Blakie}(2015)}]{Bisset2015a}%
  \BibitemOpen
  \bibfield  {author} {\bibinfo {author} {\bibfnamefont {R.~N.}\ \bibnamefont
  {Bisset}}\ and\ \bibinfo {author} {\bibfnamefont {P.~B.}\ \bibnamefont
  {Blakie}},\ }\bibfield  {title} {\enquote {\bibinfo {title} {Crystallization
  of a dilute atomic dipolar condensate},}\ }\href {\doibase %
  10.1103/PhysRevA.92.061603} {\bibfield  {journal} {\bibinfo  {journal} {Phys.
  Rev. A}\ }\textbf {\bibinfo {volume} {92}},\ \bibinfo {pages} {061603}
  (\bibinfo {year} {2015})}\BibitemShut {NoStop}%
\bibitem [{\citenamefont {Chomaz}\ \emph {et~al.}(2019)\citenamefont {Chomaz},
  \citenamefont {Petter}, \citenamefont {Ilzh\"ofer}, \citenamefont {Natale},
  \citenamefont {Trautmann}, \citenamefont {Politi}, \citenamefont
  {Durastante}, \citenamefont {van Bijnen}, \citenamefont {Patscheider},
  \citenamefont {Sohmen}, \citenamefont {Mark},\ and\ \citenamefont
  {Ferlaino}}]{Chomaz2019a}%
  \BibitemOpen
  \bibfield  {author} {\bibinfo {author} {\bibfnamefont {L.}~\bibnamefont
  {Chomaz}}, \bibinfo {author} {\bibfnamefont {D.}~\bibnamefont {Petter}},
  \bibinfo {author} {\bibfnamefont {P.}~\bibnamefont {Ilzh\"ofer}}, \bibinfo
  {author} {\bibfnamefont {G.}~\bibnamefont {Natale}}, \bibinfo {author}
  {\bibfnamefont {A.}~\bibnamefont {Trautmann}}, \bibinfo {author}
  {\bibfnamefont {C.}~\bibnamefont {Politi}}, \bibinfo {author} {\bibfnamefont
  {G.}~\bibnamefont {Durastante}}, \bibinfo {author} {\bibfnamefont {R.~M.~W.}\
  \bibnamefont {van Bijnen}}, \bibinfo {author} {\bibfnamefont
  {A.}~\bibnamefont {Patscheider}}, \bibinfo {author} {\bibfnamefont
  {M.}~\bibnamefont {Sohmen}}, \bibinfo {author} {\bibfnamefont {M.~J.}\
  \bibnamefont {Mark}}, \ and\ \bibinfo {author} {\bibfnamefont
  {F.}~\bibnamefont {Ferlaino}},\ }\bibfield  {title} {\enquote {\bibinfo
  {title} {Long-lived and transient supersolid behaviors in dipolar quantum
  gases},}\ }\href {\doibase 10.1103/PhysRevX.9.021012} {\bibfield  {journal}
  {\bibinfo  {journal} {Phys. Rev. X}\ }\textbf {\bibinfo {volume} {9}},\
  \bibinfo {pages} {021012} (\bibinfo {year} {2019})}\BibitemShut {NoStop}%
\bibitem [{\citenamefont {Natale}\ \emph {et~al.}(2019)\citenamefont {Natale},
  \citenamefont {van Bijnen}, \citenamefont {Patscheider}, \citenamefont
  {Petter}, \citenamefont {Mark}, \citenamefont {Chomaz},\ and\ \citenamefont
  {Ferlaino}}]{Natale2019a}%
  \BibitemOpen
  \bibfield  {author} {\bibinfo {author} {\bibfnamefont {G.}~\bibnamefont
  {Natale}}, \bibinfo {author} {\bibfnamefont {R.~M.~W.}\ \bibnamefont {van
  Bijnen}}, \bibinfo {author} {\bibfnamefont {A.}~\bibnamefont {Patscheider}},
  \bibinfo {author} {\bibfnamefont {D.}~\bibnamefont {Petter}}, \bibinfo
  {author} {\bibfnamefont {M.~J.}\ \bibnamefont {Mark}}, \bibinfo {author}
  {\bibfnamefont {L.}~\bibnamefont {Chomaz}}, \ and\ \bibinfo {author}
  {\bibfnamefont {F.}~\bibnamefont {Ferlaino}},\ }\bibfield  {title} {\enquote
  {\bibinfo {title} {Excitation spectrum of a trapped dipolar supersolid and
  its experimental evidence},}\ }\href {\doibase %
  10.1103/PhysRevLett.123.050402} {\bibfield  {journal} {\bibinfo  {journal}
  {Phys. Rev. Lett.}\ }\textbf {\bibinfo {volume} {123}},\ \bibinfo {pages}
  {050402} (\bibinfo {year} {2019})}\BibitemShut {NoStop}%
\bibitem [{\citenamefont {Sabbatini}\ \emph {et~al.}(2011)\citenamefont
  {Sabbatini}, \citenamefont {Zurek},\ and\ \citenamefont
  {Davis}}]{Sabbatini2011a}%
  \BibitemOpen
  \bibfield  {author} {\bibinfo {author} {\bibfnamefont {Jacopo}\ \bibnamefont
  {Sabbatini}}, \bibinfo {author} {\bibfnamefont {Wojciech~H.}\ \bibnamefont
  {Zurek}}, \ and\ \bibinfo {author} {\bibfnamefont {Matthew~J.}\ \bibnamefont
  {Davis}},\ }\bibfield  {title} {\enquote {\bibinfo {title} {Phase separation
  and pattern formation in a binary {Bose}-{Einstein} condensate},}\ }\href
  {\doibase 10.1103/PhysRevLett.107.230402} {\bibfield  {journal} {\bibinfo
  {journal} {Phys. Rev. Lett.}\ }\textbf {\bibinfo {volume} {107}},\ \bibinfo
  {pages} {230402} (\bibinfo {year} {2011})}\BibitemShut {NoStop}%
\bibitem [{\citenamefont {Nicklas}\ \emph {et~al.}(2015)\citenamefont
  {Nicklas}, \citenamefont {Karl}, \citenamefont {H\"ofer}, \citenamefont
  {Johnson}, \citenamefont {Muessel}, \citenamefont {Strobel}, \citenamefont
  {Tomkovi\ifmmode~\check{c}\else \v{c}\fi{}}, \citenamefont {Gasenzer},\ and\
  \citenamefont {Oberthaler}}]{Nicklas2015a}%
  \BibitemOpen
  \bibfield  {author} {\bibinfo {author} {\bibfnamefont {E.}~\bibnamefont
  {Nicklas}}, \bibinfo {author} {\bibfnamefont {M.}~\bibnamefont {Karl}},
  \bibinfo {author} {\bibfnamefont {M.}~\bibnamefont {H\"ofer}}, \bibinfo
  {author} {\bibfnamefont {A.}~\bibnamefont {Johnson}}, \bibinfo {author}
  {\bibfnamefont {W.}~\bibnamefont {Muessel}}, \bibinfo {author} {\bibfnamefont
  {H.}~\bibnamefont {Strobel}}, \bibinfo {author} {\bibfnamefont
  {J.}~\bibnamefont {Tomkovi\ifmmode~\check{c}\else \v{c}\fi{}}}, \bibinfo
  {author} {\bibfnamefont {T.}~\bibnamefont {Gasenzer}}, \ and\ \bibinfo
  {author} {\bibfnamefont {M.~K.}\ \bibnamefont {Oberthaler}},\ }\bibfield
  {title} {\enquote {\bibinfo {title} {Observation of scaling in the dynamics
  of a strongly quenched quantum gas},}\ }\href {\doibase %
  10.1103/PhysRevLett.115.245301} {\bibfield  {journal} {\bibinfo  {journal}
  {Phys. Rev. Lett.}\ }\textbf {\bibinfo {volume} {115}},\ \bibinfo {pages}
  {245301} (\bibinfo {year} {2015})}\BibitemShut {NoStop}%
\bibitem [{\citenamefont {Stienkemeier}\ and\ \citenamefont
  {Lehmann}(2006)}]{Stienkemeier2006a}%
  \BibitemOpen
  \bibfield  {author} {\bibinfo {author} {\bibfnamefont {Frank}\ \bibnamefont
  {Stienkemeier}}\ and\ \bibinfo {author} {\bibfnamefont {Kevin~K}\
  \bibnamefont {Lehmann}},\ }\bibfield  {title} {\enquote {\bibinfo {title}
  {Spectroscopy and dynamics in helium nanodroplets},}\ }\href
  {http://stacks.iop.org/0953-4075/39/i=8/a=R01} {\bibfield  {journal}
  {\bibinfo  {journal} {J.~Phys.~B}\ }\textbf {\bibinfo {volume} {39}},\
  \bibinfo {pages} {R127} (\bibinfo {year} {2006})}\BibitemShut {NoStop}%
\bibitem [{\citenamefont {{Durastante}}\ \emph {et~al.}()\citenamefont
  {{Durastante}}, \citenamefont {{Politi}}, \citenamefont {{Sohmen}},
  \citenamefont {{Ilzh{\"o}fer}}, \citenamefont {{Mark}}, \citenamefont
  {{Norcia}},\ and\ \citenamefont {{Ferlaino}}}]{Durastante2020a}%
  \BibitemOpen
  \bibfield  {author} {\bibinfo {author} {\bibfnamefont {Gianmaria}\
  \bibnamefont {{Durastante}}}, \bibinfo {author} {\bibfnamefont {Claudia}\
  \bibnamefont {{Politi}}}, \bibinfo {author} {\bibfnamefont {Maximilian}\
  \bibnamefont {{Sohmen}}}, \bibinfo {author} {\bibfnamefont {Philipp}\
  \bibnamefont {{Ilzh{\"o}fer}}}, \bibinfo {author} {\bibfnamefont
  {Manfred~J.}\ \bibnamefont {{Mark}}}, \bibinfo {author} {\bibfnamefont
  {Matthew~A.}\ \bibnamefont {{Norcia}}}, \ and\ \bibinfo {author}
  {\bibfnamefont {Francesca}\ \bibnamefont {{Ferlaino}}},\ }\bibfield  {title}
  {\enquote {\bibinfo {title} {Feshbach resonances in an erbium-dysprosium
  dipolar mixture},}\ }\href@noop {} {\ }\Eprint
  {http://arxiv.org/abs/2006.06456} {arXiv:2006.06456} \BibitemShut {NoStop}%
\end{thebibliography}

%

\end{document}